\documentclass{epl}

\title{Hitchhiking transport in quasi-one-dimensional systems}
\author{A. V. Plyukhin}
\institute{
  Department of Physics and Engineering Physics,
University of Saskatchewan\\Saskatoon, SK S7N 5E2, Canada
}

\pacs{05.40.-a}{Fluctuation phenomena, random processes, noise, and
  Brownian motion}
\pacs{73.63.-b}{Electronic transport in nanoscale materials and structures}
\pacs{72.80.Ng}{Conductivity of disordered solids}

\begin{document}

\maketitle

\begin{abstract}
In the conventional theory of hopping transport  
the positions of localized electronic states are assumed to be fixed, and
thermal fluctuations of atoms enter the theory  only through
the notion of phonons. On the other hand,
in 1D and 2D lattices, where  
fluctuations prevent formation of long-range order,
the motion of atoms has the character of 
the large scale diffusion. In this case 
the picture of static localized sites
may be inadequate. We argue that for a certain range of parameters,
hopping of charge carriers among localization sites in a network of 
1D chains is
a much slower process than diffusion of the sites themselves. 
Then the carriers move through the network 
transported along the chains by
mobile localization sites jumping occasionally between the chains. 
This mechanism may result in temperature
independent  mobility and frequency dependence similar to that
for conventional hopping. 
\end{abstract}

\section{Introduction}
Systems consisting of weakly coupled one-dimensional (1D)
chains are ubiquitous and important. 
Anisotropic organic solids,
polymers, columnar liquid crystals
are just a few examples of 
quasi 1D systems important for applications. 
Biomaterials like DNA and proteins
in many respects behave as topologically 1D systems.   
Because of structural and dynamic disorder, 
the charge transport in these materials
is often due to  
incoherent phonon-assisted hopping of charge carriers between 
localized sites.
Most theoretical models of hopping transport in quasi-1D systems
predict a strong temperature  dependence for the carrier mobility 
of the form
$\mu\sim e^{-(T_0/T)^\alpha}$ with $\alpha\le 1$ 
for variable-range hopping near the  Fermi level
and  $\alpha=2$ for Gaussian energy distribution of hopping sites.
Although  these predictions are often consistent with experiment, 
there are a growing number of experimental findings,
mostly in soft matter systems, which are
difficult to explain within conventional models of
thermally-activated electron hopping. 
One such puzzle 
is temperature-independent mobility (TIM).  
First encountered 
in anisotropic organic crystals of polyacenes~\cite{Schein},
TIM has been later 
observed in some other molecular crystals~\cite{scond},
columnar liquid crystals~\cite{LC}, and  
conjugated polymers~\cite{CP}.
Discussion of
possible mechanisms of TIM has recently caught 
interdisciplinary attention after a
very weak temperature dependence  was reported 
for electron conductivity along the DNA  double helix~\cite{DNA}.

Because of the Einstein relation $\mu=eD/k_BT$, TIM
implies that the diffusion coefficient $D$ increases linearly  
with temperature. Such behavior is difficult to explain within
conventional hopping models
with the Miller-Abrahams expression for jump rates
which depends exponentially on temperature
$W\sim\exp\left(-\Delta\epsilon/k_BT\right)$ for jumps upward in energy
$\Delta\epsilon>0$.
A non-activated temperature dependence of $W$
may under certain conditions  be responsible for TIM.
For weak electron-lattice coupling and temperature 
much higher than  the maximal phonon energy, transition rates
for optical-phonon-assisted  hopping  depend linearly on 
$T$~\cite{Emin}.
Then the diffusion coefficient $D\sim W$ 
is proportional to the temperature,
and the mobility $\mu=eD/k_BT$ is
temperature independent. 
Although this argument may be consistently 
implemented for polyacenes~\cite{Sumi},
it can hardly be a general explanation of
TIM.  In fact, in liquid crystals, DNA, and 
some organic semiconductors  TIM
is  usually observed below a certain temperature $T<T_c$, while
an activated temperature dependence develops usually
at $T>T_c$. Clearly, the above reasoning  
can not account for such a crossover.

Another feature of the systems exhibiting TIM is 
revealed by alternating current measurements.
For relatively high
frequencies the mobility typically increases  as a power of frequency,
$\mu(\omega)\sim\omega^s$ with $0<s<1$, while for low $\omega$
the mobility is frequency independent. This behavior is 
characteristic for the hopping mechanism, which therefore 
can account for the  frequency dependence but apparently inconsistent 
with TIM. 
 
It was suggested recently by many authors that the conflict may be 
resolved taking into account dynamic disorder, i.e.
temporal fluctuations of the hopping distance and/or potential barrier.
It was found that dynamic disorder can suppress the
exponential temperature dependence in a limited temperature range
~\cite{DynDisorder}.

In this Letter we consider a model of  fluctuation driven transport
 where thermal fluctuations 
not only supply carriers with energy  to jump (as in phonon-assisted
hopping) or modify hopping rate parameters 
(as in models with dynamic disorder),
but result in large scale diffusion of localization sites.
Unlike 3D crystals, where atoms
are always close to their equilibrium positions, the motion of atoms
in 1D and 2D  lattices may be to a large extent
delocalized~\cite{Peierls}. 
Let $q_i$ is a displacement 
of the $i$th atom in a 1D harmonic chain with, for instance, fixed ends.
All atoms have the same mass $m$, the harmonic force constant is $k$,
so the highest oscillation mode has the frequency $2\omega_0$,
with $\omega_0=\sqrt{k/m}$. 
Using a normal mode transformation one can show 
(see for instance~\cite{PS}) that 
for  finite temperature $T$ 
the mean
square displacement (MSD) of an atom 
is proportional to its
distance from the chain's end,
and that the  relative displacement  of two atoms
linearly increases with their 
separation,
\begin{eqnarray}
\langle q_i^2\rangle=\frac{k_BT}{m\omega_0^2}\,i,\qquad
\langle(q_i-q_{i+j})^2\rangle=\frac{k_BT}{m\omega_0^2}\,j.
\label{base}
\end{eqnarray}
Because thermal deviations from periodic
structure in one dimension is cumulative with distance,
those atoms far from 
the ends of a long chain might vibrate over distances 
substantially longer than the equilibrium lattice 
spacing~\cite{Peierls,Mori}.
In the infinite chain the  motion of atoms is unbounded and
has a diffusive character on a long time scale:
although the relative displacement of two neighboring atoms is 
small, their MSDs diverge linearly in time.
Because of  
loss of long-range order 
(Landau-Peierls instability),
long chains in many respects 
behave more like liquids than crystals~\cite{Mori,Geisel}, and
are often referred to as 1D harmonic liquids.
The lack of long-range correlations is manifested  
in diffuse x-ray and neutron scattering, which was
observed in a number of  linear-chain compounds~\cite{scat_exp}.
Well-known example is  columnar liquid crystals consisting of stacks of
plate-like molecules arranged on a regular 2D lattice.  
Spectroscopy studies reveal liquid-like correlations in the
direction along the columns. These systems, which often exhibit
TIM~\cite{LC}, 
are essentially 2D lattices of 1D liquid columns~\cite{DLC}. 
Another example is polymers. In recent experiments with DNA coils 
the large-scale Brownian motion of monomers  was 
directly measured for both semiflexible and flexible 
polymers~\cite{Krichevsky}.

Divergence of the atomic MSD in low-dimensional lattices 
is known to have important consequences
for phase transitions, but we are unaware of any study
of its implications for carrier transport. 
In this Letter we  consider   effects of
finite mobility of localized sites in a system of
parallel chains neglecting  chains flexibility,  
anharmonicity, quantum effects, etc. 
Although oversimplified, the model
demonstrates clearly that
a finite mobility of localized sites  may result in 
TIM with a crossover to activated transport at high temperature, 
and  also in frequency dependence similar to that for 
conventional electron hopping.

Let us  assume for simplicity that each atom in a chain
is associated with a localization site for charge carriers,
and that only carrier jumps between near-neighbor atoms
are important for hopping along the chain.
The properties of phonon-assisted hopping in a static lattice
depends on the type of disorder. To be specific, we shall assume
the random energy model (known also as Gaussian disorder model)
in which energy levels of  sites are
not correlated and characterized by a
Gaussian distribution 
$g(\epsilon)\sim
\exp\left(-\frac{\epsilon^2}{2\epsilon_0^2}\right)$.
For this model, 
which is believed to be relevant to variety of
organic conductors,
the diffusion coefficient for nearest-neighbor hopping $D_h$
has the form~\cite{Bas,Bara}
\begin{eqnarray}
D_h=
\frac{a^2}{2}\,\nu\, e^{-a/L}\,e^{-(\epsilon_0/k_BT)^2},
\label{Dh}
\end{eqnarray}
where $a$ is the lattice spacing, $L$ is the localization length
of the carrier wave function,
and $\nu$ is an attempt frequency. 
The exponential temperature dependence of (\ref{Dh}) suggests that
at low $T$ hopping of carriers on a static lattice 
may  be a much slower process than diffusion of atoms in the chain. 
Under these circumstances the carriers are transported along the chain
by mobile localization sites 
while transport in the perpendicular direction
is  due to carrier hopping between neighboring chains.   
We shall discuss the possibility and some properties of
this mechanism, referring to it as  hitchhiking
hopping, using two simple approximations.

\section{Independent chains}
On a time scale $t\gg \omega_0^{-1}$
the motion of a tagged atom in an isolated infinite harmonic chain is
diffusive, that  is the MSD of the atom increases linearly with time,
$\langle \Delta q^2(t)\rangle
=\langle[q(t)-q(0)]^2\rangle=2D_at$.
The velocity correlation function $C(t)=\langle v(t)v(0)\rangle$
for the atom is given by the oscillatory decaying Bessel function 
$C(t)=\frac{k_BT}{m}J_0(2\omega_0t)$,
and therefore the atomic diffusion coefficient $D_a$ is
\begin{equation}
D_a=\int_0^\infty dt \, C(t)=\frac{k_BT}{2m\omega_0},
\label{Da}
\end{equation}
where recall $\omega_0=\sqrt{k/m}$, 
$k$ is the force constant, $m$ is the mass of an atom~\cite{PS}.
The ratio of the diffusion coefficients 
for carrier hopping along a static chain (\ref{Dh}) 
to that for atomic  diffusion 
(\ref{Da})  
can be written as
\begin{equation}
\frac{D_h}{D_a}=\frac{\nu}{\omega_0}\,\,
\frac{ka^2}{\epsilon_0}\,\,
e^{-a/L}
\left\{\frac{\epsilon_0}{k_BT}e^{-(\epsilon_0/k_BT)^2}\right\}.
\label{con}
\end{equation}
The attempt frequency $\nu$ depends on many factors,
but usually is of order of the characteristic phonon frequency, so 
the first factor $\nu/\omega_0$ is of order one. 
The second factor is the ratio of interaction energy of near-neighbor
atoms in the chain $ka^2$  to the width of energy distribution of
localized states $\epsilon_0$. 
Taking for estimation $k\sim 10 \,\, N/m\sim 1\,\,eV/{\AA}^2$, 
$\epsilon_0\sim 0.1\,\,eV$, and $a\sim 10\,{\AA}$,  one gets 
for the second factor $ka^2/\epsilon_0\sim 10^3$.
This large value is typically compensated by the tunneling factor
$\exp(-a/L)$, which  is of order $10^{-3}$ for $a/L=6$. 
Then the  order of the ratio $D_h/D_a$ is governed by the 
expression in brackets which is small for $\epsilon_0/k_BT>1$.
In what follows we shall focus on the limit $D_h/D_a\ll 1$,
neglecting completely conventional hopping along the chains,  
and assuming that the only way for the carriers to move  
along the chains is be transported  
by diffusing localization sites. 

For parallel oriented chains one can expect that
transition rates for interchain hopping
do not strongly fluctuate
around a typical value $W_\perp$. Then
the diffusion 
coefficient for the carrier motion in the direction perpendicular 
to the chains is 
$D_\perp=\frac{1}{2} W_\perp b^2$ 
where $b$ is the distance between near-neighbor 
chains. 
For the direction along the chains 
the diffusion coefficient of the carriers can be estimated as 
\begin{equation}
D_\parallel=\frac{1}{2} W_\perp \langle \Delta q^2(W_\perp^{-1})\rangle,
\label{Destim}
\end{equation}
where $\langle \Delta q^2(t)\rangle$ is the atomic MSD.
Indeed, a carrier, riding on a diffusing localized center,  
travels a distance $\Delta q(W_\perp^{-1})$
before jumping to another chain. For time $t$ the number  
$W_\perp t$ of such rides occurs, and the MSD of the carrier
is 
$\langle R^2(t)\rangle =
W_\perp t\,\langle \Delta q^2(W_\perp^{-1})\rangle=2D_\parallel t$.
With
$\langle \Delta q^2(t)\rangle=2D_at$,
Eq.(\ref{Destim}) gives   
the diffusion coefficient for carriers $D_\parallel$ 
equals to that for atoms 
\begin{equation}
D_\parallel= D_a=\frac{k_BT}{2m\omega_0}.
\label{D}
\end{equation}
It does not depend on the interchain jump frequency $W_\perp$,  
and increases linearly with temperature.
The corresponding mobility
$\mu_\parallel=eD_\parallel/k_BT$
is temperature independent,
$\mu=e/2m\omega_0$.
For $\omega_0=10^{13}\, s^{-1}$ and $k=10\,N/m$ one 
gets $\mu\sim 10^{-3}\,cm^2\, V^{-1}s^{-1}$. This value 
is 2-3 orders of magnitude
less  than in molecular crystals like anthracene, but consistent with 
data  reported for columnar liquid crystals~\cite{LC} and 
certain conjugated polymers~\cite{CP}.

The above estimation implies diffusive motion of atoms in a chain, 
$\langle \Delta q^2(t)\rangle= 2D_a t$, which 
is valid for infinite chains only asymptotically, for 
$t\gg\omega_0^{-1}$.   One may wish to improve  
the formula (\ref{D}) using the {\it exact} expression for the 
atomic MSD  which can be obtained, 
integrating the velocity correlation function
$\langle \Delta q^2(t)\rangle=\int_0^t d\tau\,(t-\tau)C(\tau)$,
which gives
\begin{equation}
\langle \Delta q^2(t)\rangle=2D_a t-\frac{D_a}{\omega_0}
\,\,\varphi(2\omega_0 t)
\label{De}
\end{equation}
Here the first term describes the diffusive displacement,
while the second term is the more slowly
growing correction with 
$\varphi(x)=\int_0^x d\tau\,\tau J_0(\tau)$.
From (\ref{Destim}) and (\ref{De}) one gets
$D_\parallel=D_a+D_a x\,\varphi\left(\frac{2}{x}\right)$
where $x=W_\perp/\omega_0\ll 1$.

Let us now discuss some elementary predictions of the model 
concerning mobility in an  ac field of frequency $\omega$. 
Since hitchhiking transport is diffusive on  
the  long-time scale $t\gg W_\perp^{-1}$,
one can expect that 
the dynamical mobility $\mu(\omega)$ in the low-frequency limit
 $\omega\ll W_\perp$ has the Drude
form.
We shall concentrate here on the high-frequency domain
$W_\perp\ll\omega<\omega_0$ when interchain hopping is  negligible
and the  response of the system is determined entirely by dynamics of the
charged localized  sites (atoms with attached charge carriers)
driven by an external {\it ac} field.

The response of the chain to a local  ac driving force 
$F_{ex}=e\, Re\, \{E_0e^{i\omega t}\}$
applied to a single charged atom is convenient
to analyze writing the equation of motion of the atom 
in the form of the non-Markovian Langevin equation~\cite{PS},  
\begin{equation}
\dot{v}(t)=-2\omega_0^2\int_0^t dt' M(t-t')v(t')+f(t)+
f_{ex}(t).
\label{LE}
\end{equation}
Here $f_{ex}=F_{ex}/m$, and 
$f(t)$ stays for  the superposition of terms depending on 
the initial coordinate of the tagged atom, and also
on the initial coordinates and velocities of all other atoms in the chain.
This superposition of oscillating terms with uncorrelated phases 
behaves like a
random function with zero average and a finite correlation time. 
The memory kernel $M(t)$, which is essentially the correlation
function of $f(t)$, is given explicitly by 
$M(t)=J_1(2\omega_0 t)/\omega_0 t$. 
Using Eq. (\ref{LE}) with $f_{ex}=0$ one can easily find  
the velocity correlation function 
$C(t)=\langle v(0)v(t)\rangle=\frac{k_BT}{m}J_0(2\omega_0 t)$, the result
which we have already used above. 

Taking average of (\ref{LE}) and 
making a substitution 
$\langle v(t)\rangle=Re\{\mu(\omega)E_0e^{i\omega t}\}$
one obtains  for the stationary dynamical mobility $\mu(\omega)$
the expression
\begin{equation}
\mu(\omega)=\frac{e}{m}\,\,\frac{1}{i\omega+2\omega_0^2\tilde M(\omega)}.
\label{mu}
\end{equation} 
Here $\tilde M(\omega)$ is the Laplace-Fourier transform of the
damping kernel, $\tilde M(\omega)=\int_0^\infty dt\, e^{-i\omega t}M(t)$.
On the other hand, one can get from Eq. (\ref{LE}) the transform of 
the velocity correlation function
$\tilde C(\omega)=\frac{k_BT}{m}\{i\omega+2\omega_0^2\tilde M(\omega)\}^{-1}$.
Comparing this with (\ref{mu}) one obtains the Einstein relation
for dynamical mobility
\begin{equation}
\mu(\omega)=\frac{e}{k_BT}\tilde C(\omega).
\label{mu*}
\end{equation}
Alternatively, one can express $\mu(\omega)$ in terms of 
the Laplace transform of the MSD $\sigma(s)=\int_0^\infty
dt e^{-st}\langle \Delta q^2(t)\rangle$, 
\begin{equation}
\mu(\omega)=\frac{e}{k_BT}(i\omega)^2\sigma(i\omega),
\label{mu**}
\end{equation}
which can be obtained from (\ref{mu*}) taking into account
the relation 
$\frac{d^2}{dt^2}\langle \Delta q^2(t)\rangle=
2\langle v(t)v(0)\rangle$.


From (\ref{mu*})  one finds that
for $\omega<2\omega_0$  dynamical mobility $\mu(\omega)$
is purely real, 
\begin{equation}
\mu(\omega)=\frac{e}{m}\frac{1}{\sqrt{4\omega_0^2-\omega^2}}.
\label{mu2}
\end{equation}
For $\omega>2\omega_0$ the mobility is purely imaginary and 
the system does not absorb energy from the field.

\section{Dissipative chains}
Although the interchain interaction
is weak in quasi 1D systems, it may determine long-time dynamics
and  low-frequency response of the system. Also, in many soft matter systems
the chains are surrounded by solvent molecules which essentially 
affect dynamical properties of the system.
In what follows we shall assume that both interchain and chain-solvent
interactions
can be modeled with the Langevin approach,
i.e. by introducing in the equation of motion for a tagged $i$th atom 
the regular damping term $-\gamma\dot q_i$ and the random force
$\xi_i(t)$,
\begin{equation}
m\ddot{q_i}(t)=F_i(t)-\gamma \dot{q_i}(t)+\xi_i(t),
\label{LEF}
\end{equation}
where $F_i$ is the force from neighboring atoms. 
The random force $\xi_i(t)$ is assumed to be zero
centered, white noise, not correlated for different atoms,
\begin{equation}
\langle \xi_i(t)\rangle=0,\,\,\,\,\,
\langle \xi_i(t)\xi_k(0)\rangle=2k_BT\gamma\delta_{ik}\delta(t).
\label{statistics}
\end{equation}
The motion of a tagged atom in a chain surrounded by an external  bath
is an example of the transport mechanism  known as 
single file diffusion, when diffusing particles are constrained to move
in one direction and are not allowed to pass each other. Single file
diffusion is anomalous in the  sense that the MSD of the particles
asymptotically grows with time sublinearly
\begin{equation}
\langle \Delta q^2(t)\rangle=2F\sqrt{t}.
\label {SFD}
\end{equation}
For the overdamped harmonic chain the factor $F$ is~\cite{Imry} 
\begin{equation}
F=\frac{k_BT}{\omega_0\sqrt{\pi\gamma m}}.  
\label{F}
\end{equation} 

Hitchhiking hopping takes a place when the MSD of localized sites,
Eq.(\ref{SFD}), 
exceeds that for carrier diffusion due to hopping along the 
static chains
$2D_ht$, on a time scale $t\sim W_\perp^{-1}$. 
This requirement leads to the condition
\begin{equation}
\frac{D_h}{F}W_\perp^{-1/2}\ll 1.
\label{cond}
\end{equation}   
Suppose this inequality is satisfied, 
and a carrier riding on an atom for a time $W_\perp^{-1}$
travels a distance much longer than
the chain spacing $a$, passing many localization sites on neighbor chains.
Then one can expect that the thermal factor for dominating 
interchain jumps is of order of one, and 
the typical transition rate for interchain hopping
can be estimated as $W_\perp=\nu e^{-b/L}$ where $b$ is the distance between 
neighboring chains. 
Then the condition (\ref{cond})
can be written in the form
\begin{equation}
\frac{ka^2}{\epsilon_0}
\left(\frac{\gamma\nu}{m\omega_0^2}\right)^{1/2}
e^{(b-2a)/2L}
\left\{\frac{\epsilon_0}{k_BT}e^{-(\epsilon_0/k_BT)^2}\right\}\ll 1.
\label{cond2}
\end{equation} 
Since the first three factors here may be large, this condition
is a stronger constraint than that for the independent chains, 
Eq.(\ref{con}). Yet it may be satisfied
for sufficiently low temperature. Then 
the diffusion coefficient for  hitchhiking
transport, according to (\ref{Destim}) and (\ref{SFD}),
is given by
\begin{equation}
D_\parallel=
\frac{k_BT}{\omega_0}\sqrt{\frac{W_\perp}{\pi\gamma m}}.
\label{Dpar}
\end{equation} 
In contrast to the approximation of noninteracting chains,
$D_\parallel$ depends on
the interchain jumping rate $W_\perp$ which may result
in a weak temperature dependence of the mobility.
According to Eq. (\ref{mu**}), the dynamical mobility $\mu(\omega)$
is frequently independent for $\omega<W_\perp$, while for higher frequency
it increases as $\omega^{1/2}$ as a result of subdiffusive motion (\ref{SFD})
at $t<W^{-1}_\perp$. 

These results may be improved taking into account that
the single file diffusion law  (\ref{SFD}) holds only for sufficiently long
time $t\gg t_c=\alpha^{-1}$, $\alpha=2m\omega_0^2/\gamma$,
while the normal diffusion
with $D_a=k_BT/\gamma$ 
takes a place for $t<t_c$.
From Eq. (\ref{LEF}) one can obtain the {\it exact} expression for the 
MSD of a tagged atom, which is valid
for the whole time scale. 
For the overdamped chain,
$\gamma/m\gg\omega_0$,
it has the form~\cite{Imry}
\begin{equation}
\langle \Delta q^2(t)\rangle=\frac{2k_BT}{\gamma}\,te^{-\alpha t}
\Big\{I_0(\alpha t)+I_1(\alpha t)\Big\},
\label{MSD}
\end{equation}
where $I_n(x)$ are the modified
Bessel functions. 
Substitution of this expression into Eq.(\ref{Destim}) gives
the general expression for $D_\parallel$  which reduces 
to the form (\ref{Dpar}) for $W_\perp t_c \ll 1$, while
in the opposite limit $W_\perp t_c \gg 1$ it gives
$D_\parallel=D_a=k_BT/\gamma$.   
The dynamical mobility $\mu(\omega)$ can be estimated
taking the Laplace transform of the MSD (\ref{MSD}) and
then using Eq.(\ref{mu**}).  However, separation of 
real and imaginary parts is easier expressing $\mu(\omega)$ in
terms of the velocity correlation function $C(t)$.
Using the normal mode analysis similar to that in Ref.~\cite{Imry},
we obtained for overdamped motion 
\begin{equation}
C(t)=\frac{k_BT}{m}e^{-\frac{\gamma}{m} t}+
\frac{k_BT}{\gamma}\,\,\frac{d}{dt}\left\{
e^{-\alpha t}\,\,I_0(\alpha t)\right\}.
\end{equation}
Then using Eq. (\ref{mu*})  it is straightforward to get
an explicit but rather lengthy expression for $\mu(\omega)$.
For $\omega\ll\alpha$ it reduces to a simple function
with both real and imaginary parts increasing as $\sqrt{\omega}$,
\begin{equation}
\mu(\omega)=\frac{e}{2\omega_0}\sqrt{\frac{\omega}{2m\gamma}}(1-i),
\end{equation}
while in the  high frequency   domain $\omega\gg \alpha$ the mobility becomes 
frequency independent.

In conclusion, the model of hitchhiking transport 
gives a simple explanation of TIM and, for dissipative chains,
leads to a power frequency dependence.
Such behavior may be
expected for sufficiently low temperature, while for high
temperature the transport is dominated by conventional hopping,
and therefore, a crossover to a strong  temperature dependence
must occur. 
These predictions are 
consistent with experimental results discussed in the beginning 
of the Letter. 
One may speculate that the model 
may also be relevant  to systems where the nature 
of the large-scale motion of localized sites is different than that
for the simple model presented here.
For example,  conformational motion in complex biomolecules
may result in the subdiffusive MSD of molecular units
$\langle q^2(t)\rangle = Ft^{\beta}$, with $0<\beta<1$
and  $F\sim T$~\cite{Yang}. 
For this case
Eq.(\ref{mu**}) predicts temperature-independent hitchhiking mobility
with frequency dependence $\omega^{1-\beta}$.
Note that our discussion implied that the motion of atoms in a chain is 
unbounded, which is
correct strictly speaking only for an infinitely long chain.
Application of the model to finite systems may require some modification
when the atomic MSD $\langle q_i^2(t)\rangle$ on a time scale of interchain
jumps $t\sim W_\perp^{-1}$ becomes comparable to
the upper bound given by Eq.(\ref{base}).
This issue, as well as the effects of zero-point
fluctuations, will be addressed in a future publication.
Also it would be interesting to generalize 
this study in the spirit of Ajdary's model of
transport by active filaments~\cite{Ajdari} where carriers may spent
a finite time in  a solvent before being re-captured by chains.

\acknowledgments
I appreciate discussions with R. Bowles, M. Bradley, J. Schofield,
and C. Soteros. 
The work was supported by a grant from the NSERC.

\end{document}